\documentclass[conference]{IEEEtran}
\usepackage[utf8]{inputenc} 
\usepackage{cite}
\usepackage{amsmath,amssymb,amsfonts}
\usepackage{algorithmic}
\usepackage{graphicx}
\usepackage{textcomp}
\usepackage{float}
\usepackage{xcolor}

\def\BibTeX{{\rm B\kern-.05em{\sc i\kern-.025em b}\kern-.08em
    T\kern-.1667em\lower.7ex\hbox{E}\kern-.125emX}}

\begin{document}

\title{Detection of blue whale vocalisations using a temporal-domain convolutional neural network}

\author{
\IEEEauthorblockN{Bryan Sagredo$^1$\qquad Sonia Espa\~nol-Jim\'enez$^2$ \qquad Felipe Tobar$^{1,3}$}
\vspace{1em}
\IEEEauthorblockA{
\textit{$^1$Center for Mathematical Modeling, Universidad de Chile}\\
\textit{$^2$Fundaci\'on Meri}\\
\textit{$^3$Initiative for Data \& Artificial Intelligence, Universidad de Chile}
}
}

\maketitle

\begin{abstract}
We present a framework for detecting blue whale vocalisations from acoustic submarine recordings. The proposed methodology comprises three stages: i) a preprocessing step where the audio recordings are  conditioned through normalisation, filtering, and denoising; ii) a label-propagation mechanism to ensure the consistency of the annotations of the whale vocalisations, and iii) a convolutional neural network that receives audio samples. Based on 34 real-world submarine recordings  (28 for training and 6 for testing) we obtained promising performance indicators including an Accuracy of  85.4\% and a Recall of  93.5\%. Furthermore, even for the cases where our detector did not match the ground-truth labels, a visual inspection validates the ability of our approach to detect possible parts of whale calls unlabelled as such due to not being complete calls.
\end{abstract}
\begin{IEEEkeywords}
whale vocalisation, audio analysis, label propagation, convolutional neural networks, whale detection
\end{IEEEkeywords}

\section{Introduction}

The blue whale (\emph{Balaeonoptera Musculus}) is known to be the largest animal to have ever lived on Earth, with different species covering all oceans except the Arctic. Historically, blue whales have been subject to a number of threats, both man-made and natural, including being hunted by the whaling industry, colliding with large ships, and becoming entangled in fishing nets. As a matter of fact, blue whales are currently an endangered species under the Endangered Species Act and have been close to extinction during the 20th century~\cite{thomas2016status}.

To monitor the activities and behaviour of blue whales, a standard methodology is that of passive acoustic monitoring \cite{PAM}, which is particularly focused on the acquisition of communication data. In particular, analysing the sound environment of whales enables us to understand their response to other sources of sound including environmental noise and anthropogenic noise such as nearby ships~\cite{chileanwhales}.

Our case study focuses on the Chilean blue whales from the Chiloé-Corcovado region, Chile \cite{chileanwhales2,chileanwhales3}. This area is a relevant feeding ground that hosts approximately 300 blue whales, a relatively large population discovered in 1993 which has been the subject of a number of researchers lately. The recordings considered in this work were obtained by Marine Autonomous Recording Units (MARUs): buoys fixed to the seafloor with high sensitivity, both in amplitude and frequency, that are capable of recording blue whale vocalisations kilometres away. MARUs have been instrumental for the development of bioacoustic studies, as they can be deployed to cover a larger region than cable-based hydrophones. 

\section{Proposed stages: preprocessing, labelling \& supervised learning}

Previous work has focused on unsupervised detection of blue whale calls based on Bayesian Gaussian Mixture Models, and using temporal and spectral features of audio segments~\cite{chilecon2017}. In our work we consider the data in the temporal domain, that is, direct audio recordings rather than the usually-considered Spectrogram (or other Fourier/wavelet frequency domain features). 

The first stage of our proposed method is to preprocess the acoustic data, where we perform temporal segmentation, spectral denoising, bandpass filtering, and normalisation; all of these aim to condition and unify the recordings for later processing.  

Then, the data considered enter into a stage called \textit{label propagation}. The recordings considered in our study include expert labels identifying \textbf{complete whale vocalisations}: positive labels denote the presence of a complete call while negative labels the lack of it. These labels are intended for the isolation of whale calls from the recording for further analysis and, in their original format, they are not suitable for building automatic detection systems. This is because the labels do not account for partial calls, meaning that two short segments that are similar and part of a call are labelled differently if one of them is part of the complete call and the other one is not. In practice, having this label inconsistency will hinder the training of automatic recognition models, especially those based on neural networks as is our case. Therefore, in our study, we verify negative labels in each short audio segment and compare them against samples with positive labels. Then, positive labels are \textit{propagated} towards similar, negative-labelled, segments with the aim to build a consistent training set for calibrating automatic-detection algorithms. 

The third part of our detector relies on a temporal-domain convolutional neural network (CNNs)~\cite[Ch.~9]{DL_book}, that is, a CNN that receives audio recordings (after the preprocessing mentioned above). This is in sharp contrast with the standard practice in the literature, which for general applications of audio processing using neural networks, considers the input to the network as the Periodogram, Spectrogram, or another spectral representation of the data. We believe that the use of temporal representation of the recording facilitates the interpretation of the detection system, this is because since the first layers of the trained network determine the (temporal) waveforms that bioacoustic experts visually identify across the recording in order to detect whale calls. 

\section{Methodology}
\subsection{Dataset}
Our study considered 34 15-minute-long audio recordings totalling 8.5 hours of labelled data, collected between February and May 2012. The sampling rate of the recording was 2000 [Hz]. Within each  recording, every data point was annotated by an expert with the label \textsc{positive} (i.e., containing a complete whale call) or \textsc{negative} (i.e., \textbf{not} containing a complete whale call). Each recording contained between 4 and 30 whale calls, where a whale or call, or \textit{song}, consists of a set of whale vocalisations. In the data considered, calls are identified by a sequence of consecutive \textsc{positive} labels.

\subsection{Preprocessing}
\label{sec:preproc} 

The data preprocessing stages applied to the recordings prior to feeding them to the CNN are described as follows.

\begin{itemize}
\item \textbf{Audio segmentation}: We first split each audio file into \textit{windows} of 2.5 [s], with an overlap of 1 [s]. The size of the window had to be large enough to contain the stationary part of a whale vocalisation and short enough in order not to contain non-stationary data; here, we relied on the assumption that whale calls, just like human speech, are piece-wise stationary. Additionally, we assigned a single label for each window rather than one label for each data point as in the original annotations: we assigned a \textsc{positive} label to a window if and only if all the data points in it were labelled \textsc{positive}. The 2000 [Hz] sampling rate makes each 2.5 [s] window 5000 frames long.

\item \textbf{Power spectral floor denoising}: We applied a denoising filter to each audio recording in order to increase the robustness of the proposed method in the presence of noise. The method consisted in removing the \emph{power spectral floor} of the recording, which is the power spectrum (that is, the squared magnitude of the Fourier transform) of a typical\footnote{We obtain this typical noise window as the 25th window percentile, by ordering of their maximum power spectrum value.} noise window. The suppression acts on the Fourier transform of each window, controlled by a sigmoid activation function. Specifically
\begin{equation}
F_{\text{denoised}}(y)=F(y) \cdot \sigma \Big(\alpha \, \frac{P(y) - \beta\cdot P_{\text{floor}}(y)}{\beta \cdot P_{\text{floor}}(y)} \Big)\;,
\end{equation}
where:
\begin{itemize}
    \item $F(y)$ is the Fourier transform of the recording $y$,
    \item $P(y) \equiv || F(y) ||^2$ the Fourier power spectrum of $y$,
    \item $P_{\text{floor}}$ denotes power spectral floor.
\end{itemize}
The parameters $\alpha$ and $\beta$ control the smoothness of the sigmoid function and the scaling of the power spectral floor, respectively. The values used in our experiments  were $\alpha=2.5$ and $\beta=50$.

\item \textbf{Bandpass filtering}: Next, we applied a bandpass filter to each audio recording in order to attenuate frequency components that are outside the frequency region of the characteristic blue whale calls, which is roughly between 20 and 200 [Hz]. This step was crucial for the accuracy of detection for the classifier.

\item \textbf{Window normalisation}: A relevant step needed to cater for the dissimilar volumes (or amplitudes) of different recordings was to normalise each audio window. This allowed us to improve the detection of subtle (low amplitude) vocalisations. The normalisation consisted of simply subtracting the mean amplitude and dividing by its standard deviation.

\end{itemize}

\subsection{Label propagation}
The main challenge in our experiments was that the expert annotator only declared a segment of the recording as \textsc{positive} if a complete call is present. This was because the original annotations aimed to identify and study the whale calls rather than detecting them, however, in our analysis, short segments (windows) are the minimal data object to be processed. Therefore, two rather similar windows can have different labels in the case where one of them is part of a vocalisation and the other is not. 

To alleviate this situation, we assigned labels to partial calls that were labelled as \textsc{negative} by propagating the labels using a similarity criterion. For each recording, we computed the inner product between \textsc{positive} and \textsc{negative} windows. Then, recalling that windows are normalised, whenever this inner product became close to 1 we changed the \textsc{negative} label to a \textsc{positive} one, arguing that the label was \textsc{negative} because the call was incomplete, yet the recording still had vocalisation activity. The threshold considered for label assignment was 0.95. This method can only be performed in a single, continuous recording, where background noise conditions are similar for every window contained in it. Plus, it is more likely to detect a single whale emitting a series of calls similar to each other.

\subsection{An audio-fed convolutional neural network}

\begin{figure}[t]
    \centering
    \includegraphics[width=0.45\textwidth]{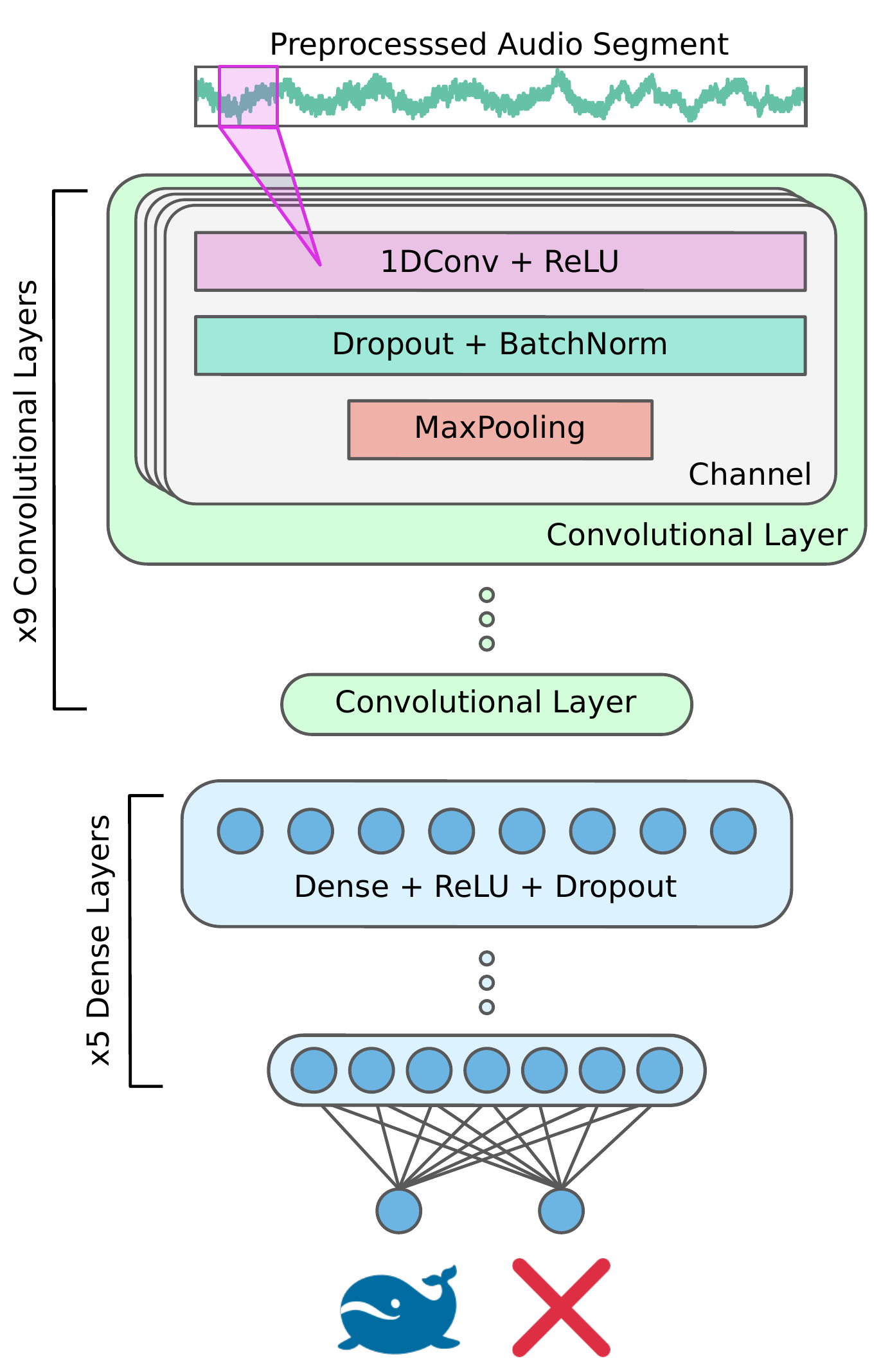}
    \caption{Diagram of the neural architecture designed for the acoustic  detection of blue whales. Convolutional layers have a filter size 8, padding size 6 and dropout probability of 0.01, the number of channels for each of the 9 convolutional layers are \{4, 8, 16, \ldots , 16\}. Max-pooling layers have a kernel of size 2.  Dense layers have a dropout probability of 0.1, and the width of each fully-connected layer was \{160, 96, 48, 32, 16\}.}
    \label{fig:CNN}
\end{figure}

The CNN architecture designed for our experiment takes audio (not frequency) samples resulting from the preprocessing stages described above and it is illustrated in Fig.~\ref{fig:CNN}. The input to the network is 2.5-second audio recordings which enter into a series of 9 convolutional layers followed by 5 fully-connected layers. The output layer is composed of two neurons, which are chosen so that the network delivers a distribution over two possible alternatives: \textsc{positive} and \textsc{negative}. Furthermore, training of the proposed networks was based on the cross-entropy loss and the Adam optimiser \cite{kingma2015adam}. 

A relevant aspect of our neural architecture was the use of two different regularisation strategies to prevent overfitting: we used dropout and $L_2$ regularisation (or \textit{weight decay} \cite{weight_decay}) with a factor of 0.001. Additionally, we considered batch-normalisation in each convolutional layer, which increases the robustness of the detector when using large batch sizes. Fig.~\ref{fig:CNN} provides more details about the used CNN.

\section{Results}

\subsection{Experimental setup}
Out of the 34 available audio files, we considered 28 files for training and validation, these were randomly divided into segments where 80\% of the data were used for training and the remaining 20\% for validation. The remaining 6 files, which comprised 3595 windows, were used for testing---label propagation was not applied to the test data. All the experiments were executed on an Intel Xeon (E5-1620v3) with 32GB of RAM and a GTX1080 GPU running Ubuntu 18.04; we used Tensorflow \cite{tensorflow2015-whitepaper} for GPU-accelerated training. The training process took roughly only two minutes.

\subsection{Performance indicators and behaviour of the detector}

\begin{figure}[t]
    \centering
    \includegraphics[trim={1.75cm 0cm 1.75cm 1cm},clip, width=0.45\textwidth]{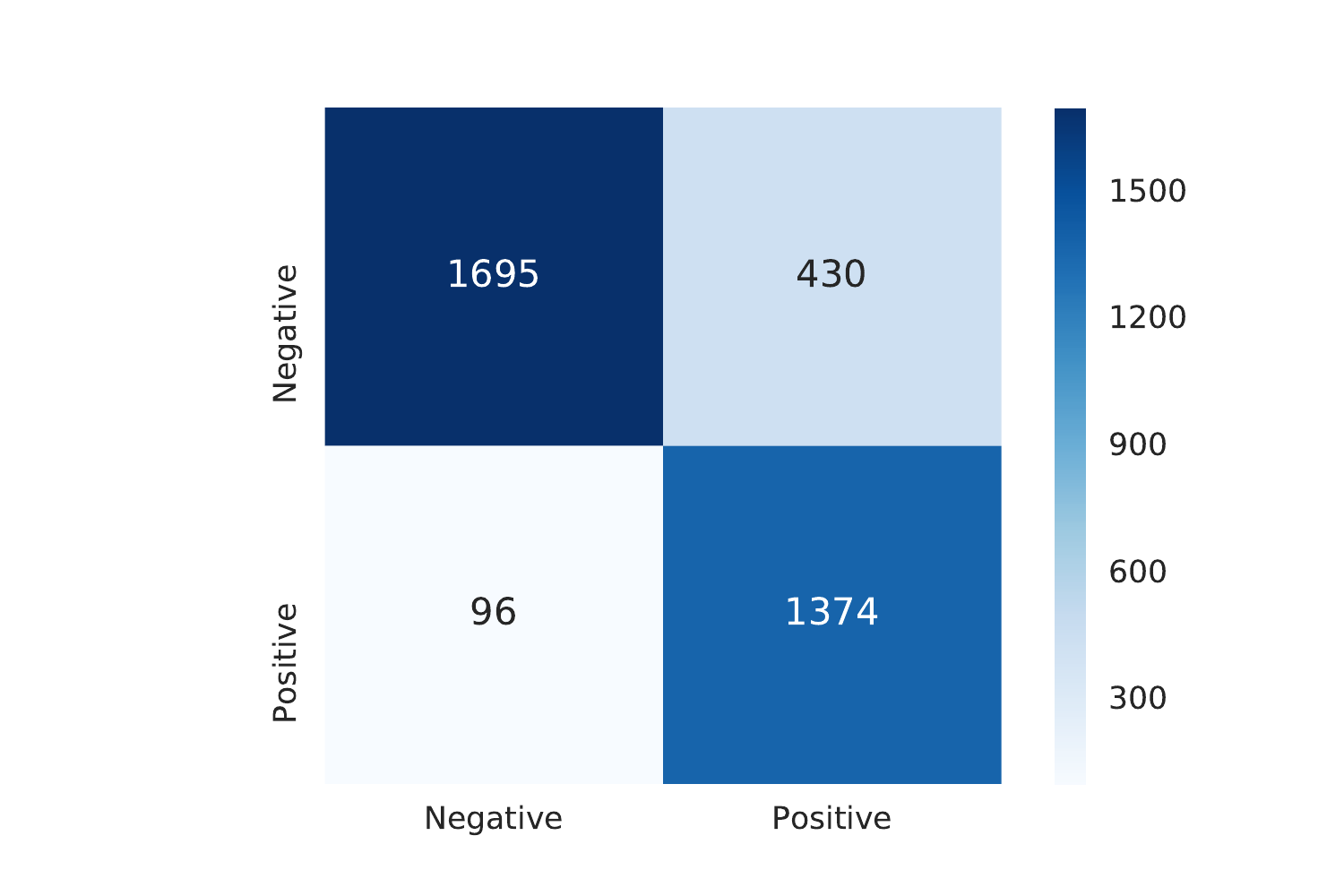}
    \caption{Confusion matrix of the 3595 test windows. On the vertical axis is the control labelling and, on the horizontal axis, the predicted labelling.}
    \label{fig:conf}
\end{figure}

 Fig.~\ref{fig:conf} shows the confusion matrix of our experiment, from where we can extract the performance indicators in Table \ref{tab:ind}.\footnote{
Recall that the indicators in Table \ref{tab:ind} are computed by
\begin{align*}
\text{Accuracy}  &= (tp+tn)/(tp+tn+fp+fn)\\
\text{Recall}    &= tp/(tp+fn)\\
\text{Precision} &= tp/(tp+fp),
\end{align*}
where $tp$, $tn$, $fp$ and $fn$ are true positives, true negatives, false positives and false negatives, respectively.} 
First, let us notice that the high Recall validates the ability of the proposed method to find vocalisation samples, since more than 93\% of all (\textsc{positive}) labelled vocalisations were identified by our method. 

On the contrary, the exhibited Precision, given by the ratio between the true vocalisations and those detected by our method, is lower and sitting at 76.2\%. However, this can be explained by the fact that expert annotations only assign \textsc{positive} labels to audio segments that contain a \textbf{complete vocalisation}, and not parts of it; since recall that expert labelling is intended to build a dictionary of whale vocalisations and not partial calls. Our detector, on the contrary, aims to detect parts of the vocalisation as well, and to do so it operates on short 2.5-second windows which are much shorter than a full whale vocalisation. Therefore, we conjecture that the \textsc{positive} predictions of the proposed method are parts of true calls that were not expert labelled. 

\begin{table}[t]
\centering
\normalsize
\caption{Performance indicators for the blue whale detector}
\label{tab:ind}
\begin{tabular}{lll}
Accuracy & Recall & Precision \\ \hline
85.4\%   & 93.5\% & 76.2\%   
\end{tabular}
\end{table}

\begin{figure}[t]
    \centering
    \includegraphics[trim={0.3cm 1cm 1.5cm 1.3cm},clip,width=0.475\textwidth]{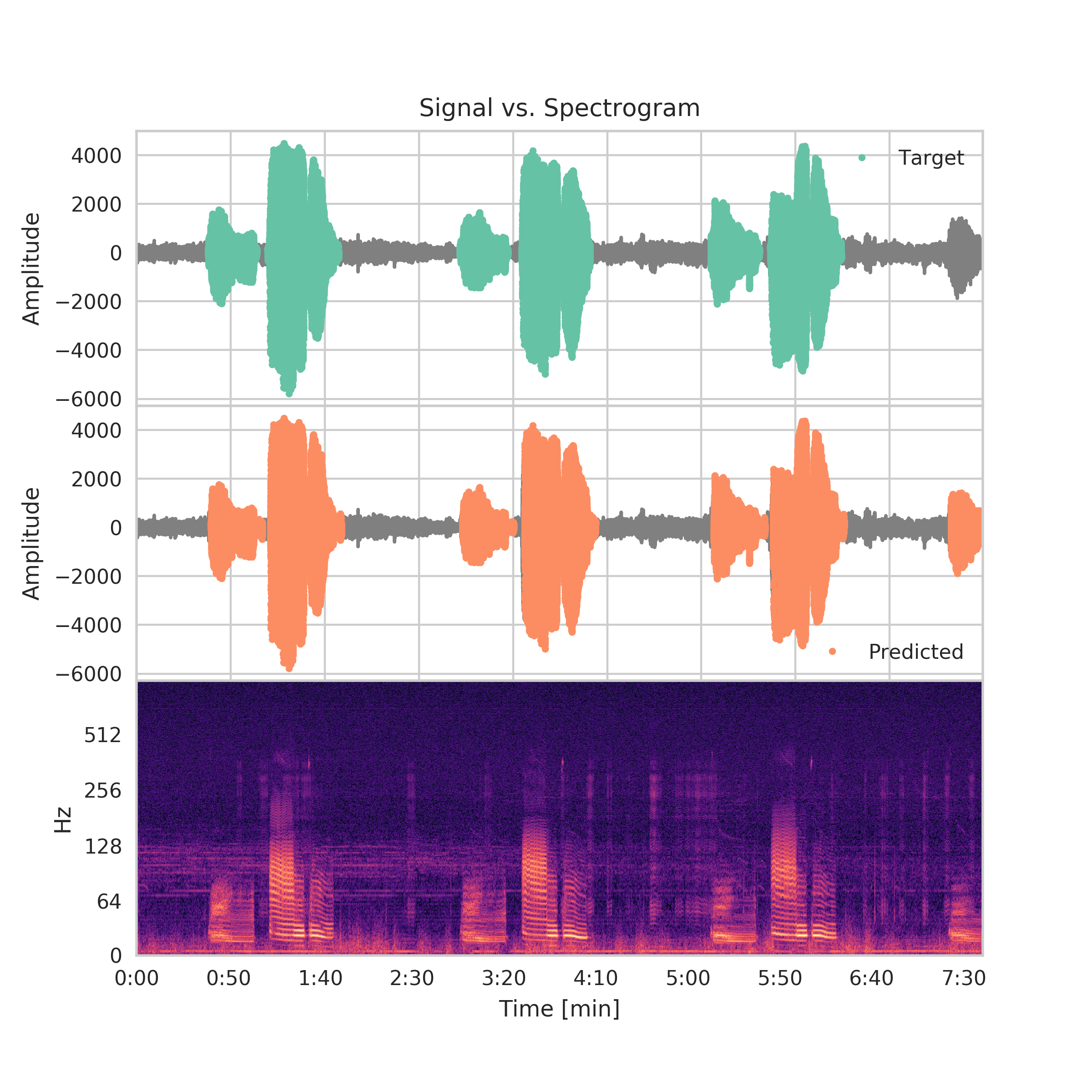}
    \caption{First test audio segment: target labels (top), predicted labels (middle) and Spectrogram (bottom). Observe that the predicted labels identify the beginning of a vocalisation.}
    \label{fig:results1}
\end{figure}

\begin{figure}[t]
    \centering
    \includegraphics[trim={0.3cm 1cm 1.5cm 1.3cm},clip,width=0.475\textwidth]{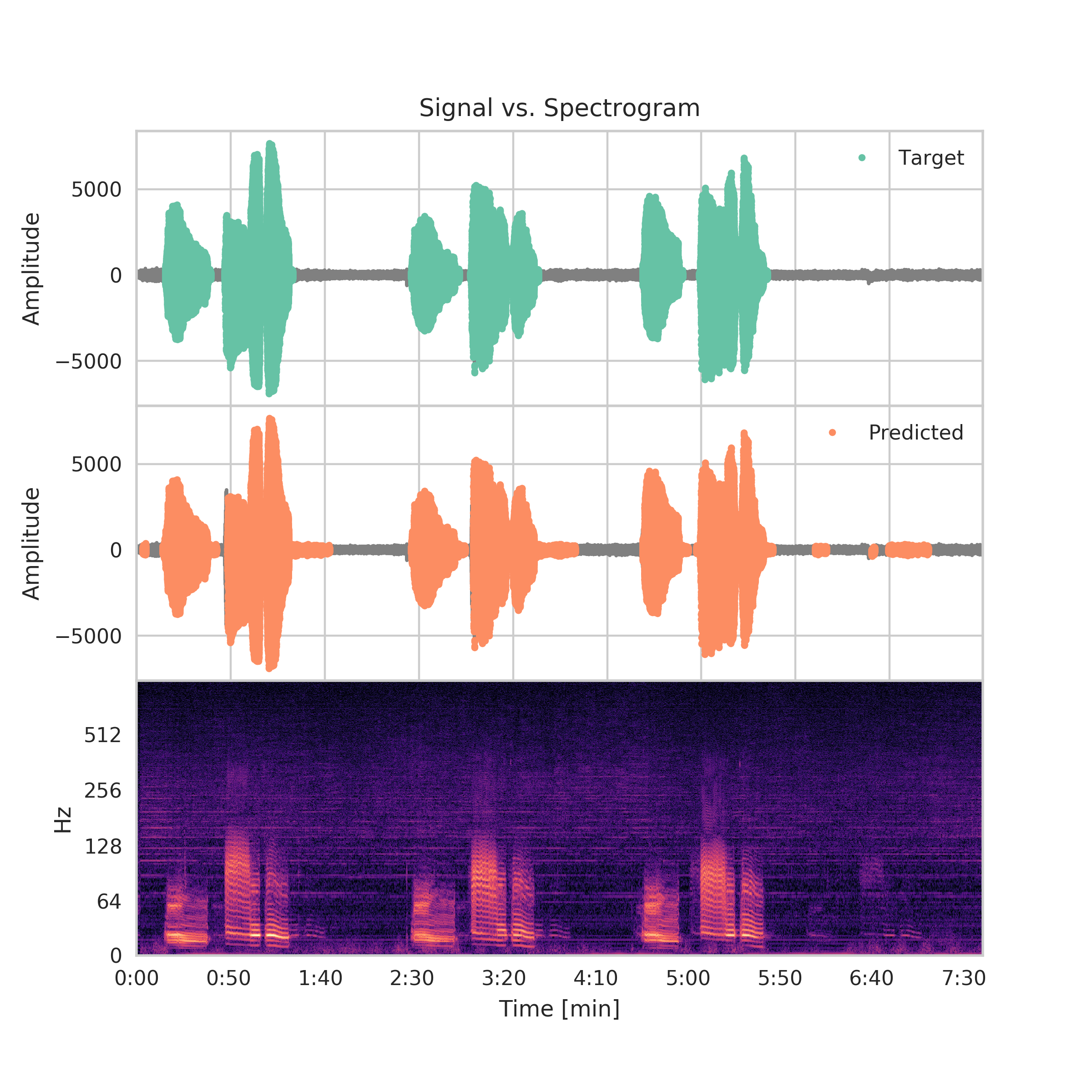}
    \caption{Second test audio segment. Observe the subtle vocalisation in the 6:40 mark that is not labelled as a vocalisation but identified by the proposed detector.}
    \label{fig:results2}
\end{figure}

\begin{figure}[t]
    \centering
    \includegraphics[trim={0.3cm 1cm 1.5cm 1.3cm},clip,width=0.475\textwidth]{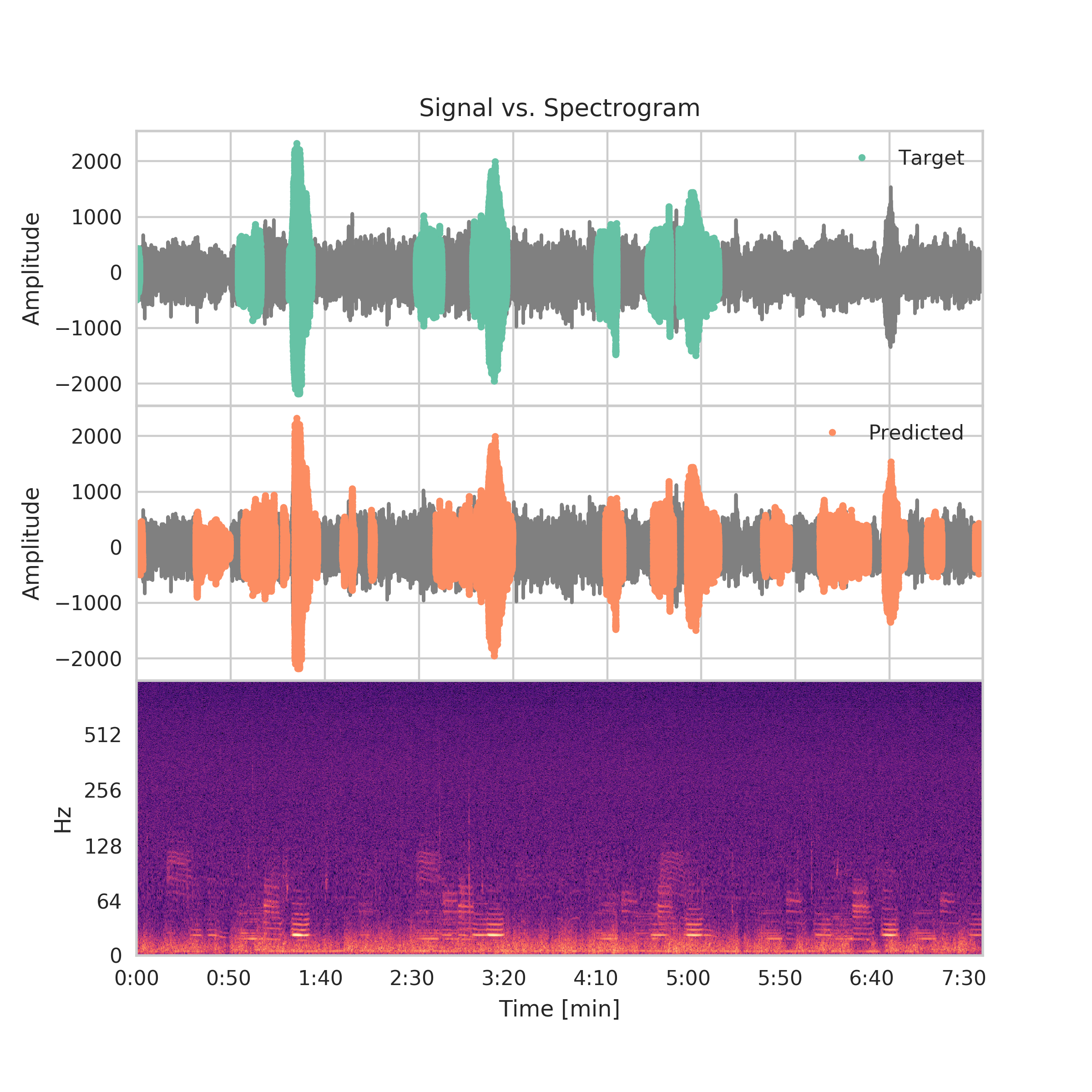}
    \caption{Third test audio segment. The predicted labels include the target ones and might identify additional calls, however, due to the low amplitude of the signals is hard to discriminate the call by visual inspection of the Spectrogram.}
    \label{fig:results3}
\end{figure}

In order to support our conjecture and also to explore the behaviour of the proposed method, we visually analyse the target and predicted labels for three audio segments of 7.5 minutes alongside their Spectrogram. These segments are presented in Figs.~\ref{fig:results1}, \ref{fig:results2} and \ref{fig:results3}, each of which shows the audio recording with the true labels in green (top), predicted labels in orange (middle) and the Spectrogram (bottom). First, observe from Fig.~\ref{fig:results1} how the ground truth vocalisations are clearly present in the Spectrogram, where the parts (phonemes) of the call are well identified. In this segment, all target labels are recovered by the proposed detector, however, notice that at the end of the recording there is a positive-predicted label not in the target (test) set. This behaviour certainly contributes to a decrease in the detector's precision, however, it can be seen from the Spectrogram that there is, in fact, the beginning of a whale call, meaning that the detector might be identifying a short vocalisation.

Fig.~\ref{fig:results2} exhibits another interesting behaviour of our detector. Notice that once again, there are additional predicted labels with respect to the target labels, this time however they correspond to subtle calls as evidenced in the Spectrogram, in particular for the region [6:40-7:30] and after the main vocalisations in 1:40 and 4:10. The detector was able to detect this low energy activity, which perhaps corresponds to vocalisation produced far from the recording unit, due to the denoising and window normalisation stages described in Sec.~\ref{sec:preproc}. 

Fig.~\ref{fig:results3} shows the behaviour of the detector for low signal-to-noise ratio, where the vocalisations are immersed in background noise. In line with the previous two test segments, the detector identified most target labels as well as other sections that might be vocalisations. In this case, however, it is less clear whether all predicted labels are in fact whale calls due to the low power of the calls.

\section{Conclusions and outlook}

We have presented a hybrid approach to blue whale call detection, which consists of a preprocessing stage inspired by classic signal processing techniques followed by a deep convolutional network that analyses submarine audio recordings in the original domain where they reside: the temporal domain. The proposed architecture has shown to be effective in the detection of the calls and even when the predicted labels do not match the expert annotations, visual inspection revealed the consistency of the proposed method. We hope that these results are useful to the bioacoustic community in order to i) complement their toolbox for analysing submarine audio, ii) reduce the burden of human analysis of audio recordings which is usually done by visual inspection of the Periodogram/Spectrogram, and iii) contribute to the greater goal of studying the fundamentals of whale communication.

The first direction of future work stemming from this study is directed towards understanding whether the proposed methodology does in fact detect true whale calls that are not labelled as such in the database. Also, by a deeper study of the CNN's activations, we aim to inspect if it is possible to identify phonemes, thus contributing to a better understanding of the underlying communication mechanism of whales.  

Beyond the proposed neural architecture, we will also consider extending our analysis to the use of recurrent neural networks (specifically, a bi-lateral long short-term memory network) to incorporate non-stationary features of the whale songs. Note, however, that training recurrent nets are more computationally expensive and might, perhaps, require the use of shorter windows.

\section*{Acknowledgements}
This work was a collaboration between the Center for Mathematical Modeling and Fundación MERI. The dataset was provided by Fundación MERI. We would like to thank Alejandro Cuevas for fruitful discussions on the data analysis methods used. This work was funded by Fondecyt-Regular 1210606, ANID-AFB170001 (CMM) and ANID-FB0008 (AC3E).

\bibliographystyle{IEEEtran}
\bibliography{references}

\end{document}